# Evidence of Low Dimensional Chaos in Glow Curves of Thermoluminescence

Elio Conte [1,2] , Joseph P. Zbilut [3]

[1] Department of Pharmacology and Human Physiology – TIRES – Center for Innovative Technologies for Signal Detection and Processing, University of Bari- Italy;
[2] School of Advanced International Studies for Applied Theoretical and Non Linear Methodologies of Physics, Bari, Italy;
[3] Department of Physiology, Rush Medical College, Rush University, Chicago, Illinois, 60612 USA

Abstract : Electron trapping following exposition to ionising radiations and consequent electron release during variation of temperature in solids represent processes happening at the quantum microphysical level. The interesting feature of the thermally stimulated process, that in fact deserves further investigation, is that the dynamic of electrons release during ,variation of the temperature, here examined through the so called thermoluminescent Glow Curve, evidences chaotic and fractal regimes. Phase space reconstruction, Correlation Dimension, largest Lyapunov exponent , Recurrence Quantification Analysis(RQA) and fractal dimension analysis, developed by calculation of Hurst exponent, are performed on three samples. The results unequivocally fix that Glow Curves respond to a chaotic regime. RQA supports such results revealing the inner structure of Glow Curve signals in relation to their properties of recurrence, determinism and intermittency signed from laminarity as well as chaos-chaos and chaos order transitions.

*General Theory of Thermoluminescence.* It is well known that in thermoluminescence (TL) dating, a TL curve is obtained when a solid sample, taken as example from a pottery, is heated after that it has remained exposed to natural irradiations ($\alpha, \beta, \gamma$) for a certain number of years. The information is stored in the form of trapped electrons, and the energy absorbed by the sample is emitted during heating as light in the form of a glow curve. There are electrons which have been ionized by nuclear radiation and which have diffused into the vicinity of a defect in the lattice that is attractive to electrons, such as a negative-ion vacancy, and have become trapped there. The nuclear radiation is from radioelements in the sample and in its surroundings; but there is also a small contribution from cosmic rays. The more prolonged the exposure to ionizing radiation the greater the number of trapped electrons, which hence increases with the years that have elapsed since the last event at which the traps were emptied.

This setting of the clock to zero is the event dated and it can be due to the agency of heat, as with pottery, or of light, as with geological sediment.

A measure of the number of trapped electrons is obtained by stimulation by heat in the case of TL . The stimulation causes the eviction of electrons from their traps whereupon they diffuse around the crystal until some form of recombination centre is found, such as a defect activated by being charged with a hole.

In conclusion, thermoluminescence is the emission of light happening when electrons (or holes) are released from the traps and return to stable states; the escape probability is greatly increased by raising temperature. If TL emission is detected and plotted as a function of time during TL temperature linear growing, a curve is obtained named Glow Curve (GC) exhibiting several peaks, eventually not resolved, corresponding to the various energies of the emptied traps. The amplitude of each peak or the subtended area are approximately proportional to the trap population and consequently to the so called absorbed dose. The population probabilities of the various traps are different for different radiation kinds, and therefore the Glow Curve shape is different too.

Quartz is one among the different materials used for thermoluminescence dating.

Halperin and Braner [1] developed a general theory based on the following linear equations to account for a single TL peak

$$I = -\frac{dm}{dt} = A_m m n_c \quad , \quad \frac{dn}{dt} = -sn\exp(-E/kT) - A_n(N-n)n_c \quad , \quad \frac{dn_c}{dt} = \frac{dm}{dt} - \frac{dn}{dt}$$

where $N$ is the concentration of the traps ($cm^{-3}$), $n$ is the concentration of electrons in traps ($cm^{-3}$), $n_c$ is the concentration of free electrons in the conduction band ($cm^{-3}$), $A_m$ and $A_n$ are the probabilities of recombination and of retrapping ($cm^3/\sec.^{-1}$), and $m$ is the concentration of holes in recombination centres ($cm^{-3}$). $I$ is the TL intensity, $I(T)$. $T$ is the absolute temperature and $k$ is the Boltzmann constant ($eV/K$).

The first theoretical treatment for the case of a single isolated peak was given by Randall and Wilkins in 1945 [2], who wrote the following equation

$$I = -C\frac{dn}{dt} = Csn\exp(-E/kT)$$

where $s$ is the frequency factor ($\sec^{-1}$).

A detailed exposition of the theoretical developments may be found in [3].
Consider a more recent equation giving TL intensity of a glow-peak in the TL glow-curve

$$I(T) = I_m \exp(\frac{E}{kT_m} - \frac{E}{kT})\exp\left[\frac{E}{kT_m}(\alpha(\frac{E}{kT_m})) - \frac{T}{T_m}\exp(\frac{E}{kT_m} - \frac{E}{kT})\alpha(\frac{E}{kT}))\right]$$

where $I$ is the glow peak intensity, $E(eV)$ is the activation energy and $T_m$ and $I_m$ are the temperature and the intensity of the maximum, respectively. $\alpha(x)$ is a quotient of $4^{th}$ order polynomials, $x = E/kT$ (for details, see [4]. The essential feature that we have to observe for interest in our paper is that, from the previous formulation, we have that the glow-curve is a non-linear function of $T_m$, $I_m$, and $E$ variables. Therefore, it has a sense to investigate in order to evidence the possible presence of chaotic deterministic and fractal dynamics.

*The Recurrences and the Variability of Signals in Nature.* Only few systems in Nature exhibit linearity. If a system is linear, outputs of this system will be in proportion to inputs. Instead, the greatest whole of natural systems, especially those who pertain to physics and to biological matter, possess in some cases a complexity that results in a great variedness and variability, linked to non linearity, to non stationarity, and to non predictability of their time dynamics. Traditional approaches for the analysis of the variability of signals usually included the application of usual time series methods. In time domain, statistical methods were first used to describe the amplitude distribution of signals and later, methodologies used spectral analysis methods. However, useful such methods may be, they suffer of fundamental limits. They are applied assuming linearity and stationarity of signals that actually could not exist. Consequently, such methods are unable to analyse in a proper way the irregularity present in most of signals and often it is disregarded or considered as coming from an external source, random in nature. Such irregularity may be instead at the basis of the dynamics that we intend to explore. It reveals that complex behaviours of the system are very distant from previously accepted principles. The study of this very irregular behaviour requires the introduction of new basic principles. Consequently, the employed methodologies of analysis must represent the counterpart of the discovery the origin of such irregularity that, in fact, arises from acting non linear mechanisms in Nature. Therefore, nonlinear science is becoming an emerging methodological and theoretical framework that makes up what is called the science of the complexity, often called also chaos theory. The term chaos here must not be intended as a lack of organization or order in the considered chaotic system but as a complex state of such system in which an apparent randomness of the system is evident but it is really constrained by a kind of order that is non linear. An important concept here is that of chaotic behaviour. It will be defined chaotic if trajectories issuing from points of whatever degree of proximity in the space of phase, distance themselves from one another over time in an exponential way. Consider some important features in non linear dynamics of systems.

*Embedding time series in phase space.* The notion of phase space is well known in physics. Let us admit we have a system, determined by the set of its variables. Since they are known, those values specify the state of the system at any time. We may represent one set of those values as a point in a space, with coordinates corresponding to those variables. This obtained construction of space is called phase space. The set of states of the system is represented by the set of points in the phase space. The question of interest is that we perform an analysis of the topological properties of phase space but, as counterpart, we obtain insights into the dynamic nature of the system. In experimental conditions, in some cases we are unable to measure all the variables of the system. In this case we may be able to reconstruct equally a phase space from experimental data where only one of the present variables (characterizing the whole system) is actually measured. The phase space is realized by a set of independent coordinates. Generally speaking, the attractor is the phase space set generated by a dynamical system represented by a set of difference or differential equations. In the actual case, let us take a non linear dynamical system represented by three independent variables $X(t), Y(t), Z(t)$, functions of time. The phase space set is given by the values of the variables at each time. The point $(x, y, z)$ in phase space gives the values of the three variables and thus the state of the system at each time. Usually, in physics, we plot one of the variables and its derivatives,

$$X, \frac{dX}{dt}, \frac{d^2 X}{dt^2}, \ldots \qquad (2.1)$$

on the three perpendicular axes $(x, y, z)$. The result is that we have reconstructed the phase space using only one of the three time series using also the derivatives of $X(t)$. This is a licit step since $Y(t)$ and $Z(t)$ are coupled to $X(t)$ through non linear equations. Consider that in experiments we have a fixed time sampling, $\Delta t$ (time series recorded at equal time intervals), and the time series is given in the following manner

$$X(0), X(\Delta t), X(2\Delta t), X(3\Delta t), \ldots\ldots\ldots\ldots, X(n\Delta t) \qquad (2.2)$$

We could also differentiate such values determining $dX/dt, d^2 X/dt^2$, ….. but such procedure is unprofitable. In fact, also if our data of time series should contain only very small errors in measurements, they should become more large errors during such operations. We may follow another procedure. We introduce a time lag $\tau = m\Delta t$ and consider each point in phase space, given by the following vector expression

$$[X(t), X(t+\tau), X(t+2\tau), X(t+3\tau), \ldots X(t+(N-1)\tau)] = \overline{X}_N (t) \qquad (2.3)$$

where $N$ is the selected dimension of the phase space. Note that assuming such procedure in phase space reconstruction we do not loss in generality since, as it is easy to show, the coordinates of the phase space reconstructed in this manner, using time delays, are linear combinations of the derivatives.

This procedure of reconstruction of phase space starting with the given time series is called *embedding*. This is the method presently used for reconstruction of phase space of experimental sampled time series. Takens in 1981 [5]showed that this embedding method, based on time lags, is certainly valid under some suitable conditions. The first requirement is that the considered time series must be twice differentiable. If this requirement is not satisfied, and it happens often, when the considered time series is a fractal, the fractal dimension, calculated by the embedding method, may also not result equal to the true fractal dimension of the phase space set. Still, the other feature relating Takens theorem, requires that in a realistic reconstruction of phase space, say of dimension $D$, we must embed in a space of dimension $(2D+1)$ in order to express enough dimensions. This is to avoid that the $N-$ dimensional orbits do intersect themselves in a false manner.

*The Determination of Time Lag* $\tau$. Some different procedures may be followed to determine the time lag of the given time series in the embedding method. The problem must be solved with particular care. The proper choice of the time lag is of fundamental importance because in chaotic signals the relation between the dimension of an embedding space and real phase space is strongly linked just to length chosen for time lag. The use of a too small lag may result in strong correlations among the components of the (2.3), and the local geometry of embedding results much like as a line (i.e. dimension equal to 1), and damaging image reconstruction of the chaotic attractor. As methodological praxis, it is useful to study the autocorrelation function of the given time series. Correlation decreases with time. The time lag is selected as the autocorrelation function reaches its first zero. Often another useful criteria is to take the time lag as the autocorrelation function decreases to $1/e = 0.37.$ In addition to use of autocorrelation function, it is necessary to employ the Mutual Information Content, $MI(\tau)$. Mean Mutual Information is given in the following manner [6]

$$MI(\tau) = \sum_{X(i),X(i+\tau)} P(X(i), X(i+\tau)) \log_2 \frac{P(X(i), X(i+\tau))}{P(X(i))P(X(i+\tau))} \tag{2.5}$$

The time at which the first local minimum of mutual information content is reached, may represent a good choice for time lag. Both $Au(\tau)$ and $MI(\tau)$ must be used, selecting the time lag provided for $MI(\tau)$ if $Au(\tau)$ and $MI(\tau)$ predict different results. This is preferable since $MI(\tau)$ also accounts for non linear contributions in time series.

*Embedding Theorem and False Nearest Neighbors*. As previously outlined, according to embedding theorem (see Takens theorem for details), the choice of dimension $N$ of reconstructed phase space should require a priori knowledge of the dimension $d_F$ of the original attractor with $N > 2d_F$. This is decisively unrealistic for time series of experimental data. Selecting $N$ in absence of a given criterion, it may result too small as compared to $d_F$ of the original attractor. It is possible to employ what is called the criterion of False Nearest Neighbors (FNN) in reconstructed phase space [6]. A point of data sets is said FNN when it comprises the local nearest neighbors not actually but only because the orbit is constructed in a too small embedded space determining its self-crossing. This difficulty may be overcame by adding sufficient coordinates to the embedding space. The criterion to use is to increase $N$ in a step manner until the number of the FNN goes substantially to zero. Usually, a threshold about 5% may result acceptable.

This strategy concludes our attempt to reconstruct phase space starting from an experimental time series.

*Fractality and Deterministic chaos of Time Series*. The use of non linear methods presumes that the signal under study is represented by an experimental time series relating a non linear system. Sometimes it possess some deterministic features that may be also chaotic and must be investigated by the methodology exposed in the previous sections.

Fractality refers to the features of a given stochastic time series. It shows temporal self-similarity. A time series is said self-similar if its amplitude distribution remain unchanged by a constant factor even when the sampling rate is changed. In the time domain one observes similar patterns at different time scales. In the frequency domain the basic feature of a fractal time series is its power law spectrum in the proper logarithmic scale. Fractals and chaos have many common points. When the phase space set is fractal, the system that generated the time series is chaotic. Chaotic systems can be arranged that generate a phase space set of a given fractal form. However, the systems and the processes studied by fractals and chaos are essentially different. Fractals must be considered processes in which a small section resembles the whole. The point in fractal analysis is to determine if the given experimental time series contains self-similar features. Deterministic chaos means that the output of a non linear deterministic system is so complex that in some manner mimes random behaviour. The point in deterministic chaos analysis is to investigate the given experimental time series that arises from a deterministic process and to understand in some manner the mathematical features of such process. Chaotic time series means that the corresponding system has sensitivity to initial conditions. When we speak about strange attractor this means that the attractor is fractal . It is very important to account for such properties since there are also chaotic systems that are not strange in the sense that they are exponentially sensitive to initial conditions but have not a fractal attractor. Still we have non chaotic systems that are strange in the sense that they are not sensitive to initial conditions but they have a fractal attractor. In conclusion, we must be care in considering fractals and non linear approaches since they are very different to each other. Often, instead, we are induced to erroneously mix different things with serious mistakes.

The geometry of the attractors is frequently examined by calculating the so called correlation dimension [7]. The self-similar property of the attractor is estimated by the scaling behaviour of the correlation integral

$$C_N(r) = \frac{1}{n^2} \sum_{i \neq j} \vartheta(r - \|\overline{X}_N(i) - \overline{X}_N(j)\|)$$

where $\vartheta(\cdot)$ is 1 for positive arguments and 0 for negative arguments. Fixed a sphere of radius $r$, in the reconstructed phase space $C_N(r)$ gives the normalized number of points in it. For stochastic signals the correlation integral, calculated in the $N$ – dimensional space, scales as

$$C_N(r) \approx r^N$$

For bounded signals there is a finite scaling exponent so that

$C_N(r) \approx r^d$ with $d < N$.

Correlation Dimension, usually indicated by $D_2$, is calculated as the slop of the linear behaviour of $\log r$ vs. $\log C_N(r)$. The value 1.0 is obtained in the case of a limit cycle, 2.0 instead is calculated in the case of a torus. A calculated non- integer value instead indicates that the phase space has a fractal geometry. However, in analysis of experimental time series the calculation of the correlation dimension does not offer results sensitive enough to conclude for a non-integer, fractal dimension that could be generated by a deterministic chaotic system. Stochastic signals may mimic chaos data and further time series of stationary data are always required.

*Estimation of Lyapunov Exponents* .As previously mentioned, chaotic systems show a dynamics where phase space trajectories with nearly identical initial states will, however, separate from each other at an exponentially increasing rate. This is usually called the sensitive dependence on initial conditions in chaotic deterministic systems. The spectrum of the Lyapunov exponents captures such basic feature of the dynamics of these systems. We may consider the two nearest neighboring points in phase space at time 0 and at time t. Let us consider also a direction i-th in space. Let be $\|\delta x_i(0)\|$ the distance at time 0 and $\|\delta x_i(t)\|$ the distance at time t. The Lyapunov exponent, $\lambda_i$ (direction i-th), will be calculated accounting that [8]

$$\frac{\|\delta x_i(t)\|}{\|\delta x_i(0)\|} = e^{\lambda_i t} \quad \text{for } t \to \infty$$

that is equivalent to

$$\lambda_i = \lim_{t \to \infty} \frac{1}{t} Ln \frac{\|\delta x_i(t)\|}{\|\delta x_i(0)\|}$$

It is possible to reconstruct the Lyapunov spectrum accounting for all the directions in phase space. Chaotic systems are characterized by having at least one positive Lyapunov exponent while their sum generally must be negative. Consider there is a whole spectrum of Lyapunov exponents, the number of them is equal to the number of dimensions of the phase space. If the system is conservative (i.e. there is no dissipation), a volume element of the phase space will stay the same along a trajectory. Thus the sum of all Lyapunov exponents must be zero. If the system is dissipative, the sum of Lyapunov exponents is negative.

The Lyapunov spectrum can be used also to give an estimate of the rate of entropy production and of the fractal dimension of the considered dynamical system. In particular from the knowledge of the Lyapunov spectrum it is possible to obtain the so-called Kaplan-Yorke dimension $D_{KY}$, that is defined as follows:

$$D_{KY} = k + \sum_{i=1}^{k} \frac{\lambda_i}{|\lambda_{k+1}|}$$

where $k$ is the maximum integer such that the sum of the $k$ largest exponents is still non-negative. $D_{KY}$ represents an upper bound for the information dimension of the system. Moreover, the sum of all the positive Lyapunov exponents gives an estimate of the Kolmogorov-Sinai entropy accordingly to Pesin's theorem [8] .In conclusion, the Lyapunov exponent is a measure of the rate at which nearby trajectories in phase space diverge. Chaotic orbits show at least one positive Lyapunov exponent. Instead periodic orbits all give negative Lyapunov exponents. It is of interest also the analysis of a Lyapunov exponent equal to zero. It says that we are near a bifurcation. It is common use in methodological activity to avoid to calculate the whole Lyapunov spectrum, estimating instead only the most positive one , usually refered to the largest one. A positive value is normally taken as indication that the system is chaotic. The inverse of the largest Lyapunov exponent is sometimes referred in literature as Lyapunov time, and defines the characteristic folding time. For chaotic orbits it is finite, whereas for regular orbits it will be infinite.Under the perspective of the analysis one must account that, for the calculation of Lyapunov exponents from limited experimental data of time series, various methods have been proposed [8]. These methods may be sensitive to variations in parameters, e.g., number of data points, embedding dimension, reconstructed time delay, and are usually reliable with care, except for long noise-free and stationary time series.

*The Method of Surrogate Data in Time Series* .At this stage of the present exposition, the reader will have realized that the most unfavourable snare in investigation of experimental time series, possibly chaotic, is that the methods we have at our availability, are inclined to give similar results in the case of deterministic chaotic dynamics and stochastic noise so that distinguishing deterministic chaos from noise becomes an important problem. Starting with a given experimental time series, stochastic surrogate data may be generated having the same power spectra as the original one, but having random phase relationship among the Fourier components. If any numerical procedure in studying deterministic-chaotic dynamics will produce the same results for surrogate data as well as for the original ones within a prefixed criterion, we will not reject the null hypothesis that the analyzed dynamics is determiend by a linear stochastic model rather than to be represneted by deterministic chaos. Often the method of the shuffled data is used. Data of the original time series are shuffled, and this operation preserves the probability distribution but produces generally a very different power spectrum and correlation function.

*Fractional Brownian Analysis in Time Seires*. It is well knwn that the study of stochastic processes with power-law spectra started with the celebrated paper on fractional Brownian motion (fBm) by Mandelbrot and Van Ness in 1968 [9]
Fixed the initial conditions, fBm is defined by the following equation

$$X(ht) \overset{d}{=} h^H X(t) \tag{3.1}$$

Given a self-similar fractal time series, the (3.1) establishes that the distribution remains unchanged by the factor $h^H$ even after the time scale is changed. ($\overset{d}{=}$) states that the statistical distribution function remains unchanged. *H* is called Hurst exponent, it varies as $0 < H < 1$, and it characterizes the general power – law scaling. For an additive process of Gaussian white noise, we have $H = 0.5$. *H* values greater than 0.5 indicate persistence in time series that is to say past trend persist into the future(long-range correlation). Instead, *H* values less than 0.5 indicate antipersistence and this is to say that past trends tend to reverse in the future. The fBm also exhibits power-law behaviour in the Fourier spectrum. There is a linear relationship between the log of spectral power vs. log of frequency. The inverse of the slope in the log-log plot is called the spectral exponent $\beta$ ($1/f^{\beta}$ behaviour), and it is related to *H* by the following relationship

$$H = \frac{\beta - 1}{2}.$$

*Recurrence Quantification Analysis.* Recurrence analysis was first introduced by Eckmann, Kamphorst and Ruelle in 1987 [10], A recurrence quantification analysis, always indicated by RQA, was subsequently introduced by Zbilut and Webber [11] and further enriched by the introduction of other variables by Marwan [12].An exceptional element of value of RQA is that this method has not restrictions in its applications: as example it may be applied also to non stationary time series.The first recurrence variable is the % Recurrence (%REC). %REC quantifies the percentage of recurrent points falling within the specified radius. Out of any doubt we may define it the most important variable in analysis of time series.The second recurrence variable is the % Determinism (%DET). %DET measures the proportion of recurrent points forming diagonal line structures. Diagonal line segments must have a minimum length in relation to above line parameter. Repeating or deterministic patterns are characterized by such variable. Periodic signals will give origin to long diagonal lines. Instead chaotic signals will give origin to very short diagonal lines. Stochastic signals will not determine diagonal lines unless a very high value of the radius will be selected.The third recurrence variable is the MaxLine (LMAX). It is the length of the longest diagonal line segment in the plot excluding obviously the main diagonal line of identity. This is a variable of valuable interest since it inversely scales with the most positive Lyapunov exponent previously discussed. Therefore, the shorter the maxline results, the more chaotic the signal is. It is a measure of divergence for the system. In addition, RQA may be performed by epochs, so that LMAX enables evaluation of Lyapunov exponent locally.The other important recurrence variable is entropy (ENT). It relates Shannon information entropy of all the diagonal line lengths distributed over integer bins in a hystogram. ENT may be considered a measure of the signal complexity and is given in bits/bin. For simple periodic systems all diagonal lines result of equal length and the entropy is expected going to zero.

Another decisive variable in RQA is the trend (TND). All the above methods discussed in the previous sections hold for stationary time series. This is a condition rarely meet in analysis of experimental time series and especially in the field of biological signals. RQA may be restfully applied for any kind of experimental time series including non stationary time series. This is one of the reasons to appreciate such RQA method. The trend (TND) still quantifies the degree of non stationarity of the time series under investigation. If recurrent points are homogeneously distributed across the recurrence plot, TND values will approach zero. If they are heterogeneously distributed across the recurrence plot, TND values will result different from zero. The sixth important variable in RQA, introduced by Marwan [12] is %Laminarity (%LAM). %LAM measures the percentage of recurrent points in vertical line structures rather diagonal line structures. Finally, the Trapping Time (TT) measures the average length of vertical line structures. Square areas (really a combination of vertical and diagonal lines) indicate laminar areas, possibly intermittency, suggesting transitional regimes, chaos-ordered, chaos-chaos transitions.

In conclusion, RQA may be considered at the moment the most powerful method for analysis of any kind of time series without limitations of any kind.

*Chaos analysis of Glow Curves in Thermoluminescence*

In our analysis we used three different Glow curves obtained from samples (Cal1, Cuc1, Raf1) of pottery that were examined for thermoluminescence dating. All the three samples corresponded to datations resulted about the period 1480- 1515 a. C. Pieces of 309 and 400 points of the Glow Curve data were used, ranging in temperature from 50 °C to 280°C(respectively to 310°C) in order to account for the most transient regime during the measurement of intensity $I(T)$ in correspondence of the number of trapped electrons stimulated by the heat. Samples (2 mg, fine grain method) were measured using an Harshaw 2000 TL system. In Figure 1 we give the Glow Curve of the sample (Cuc1) in the case of 309 points. The other samples showed similar Glow Curve behaviour.

The use of the autocorrelation function enabled us to estimate a time delay $\tau = 51.68$ for the case of Cal1-309 and $\tau = 84.62$ for Cal1-400. A time delay $\tau = 40.72$ for Cuc1-309 and a time delay $\tau = 54.33$ was obtained in the case of Cuc1-400. Finally, a time delay $\tau = 40.23$ was obtained for Raf1-309 and $\tau = 61.11$ was calculated for Raf1-400. The use of Mutual Information enabled us, instead, to estimate $\tau = 2$ for all the samples. Therefore, a time delay $\tau = 2$ was selected. Also the use of FNN criterion to estimate the embedding dimension gave homogeneous results. In fact, all the samples, examined by FNN with threshold value at 6%, gave as estimated embedding dimension the value $Dim = 6$. In Figure 2 we give the graph of % FNN against dimension in the case of Cuc1-309.

After phase space reconstruction of each of the given samples, we passed to estimate the dimension of the attractors. We obtained the following values: $D_2 = 2.361 \pm 0.276$ for the case of Cal1-309, $D_2 = 2.315 \pm 0.223$ for the case of Cal1-400, $D_2 = 1.863 \pm 0.145$ for the case of Cuc1-309, $D_2 = 1.900 \pm 0.246$ for the case of Cuc1-400, $D_2 = 2.013 \pm 0.254$ for the case of Raf1-309, and $D_2 = 1.973 \pm 0.258$ for Raf1-400. Results are given in Figures 3 and 4 for the case of Cuc1-309 and Cuc1-400.

Note that the use of Correlation Dimension to estimate $D_2$ furnished in all the case a non integer value of the dimension suggesting that we are in presence of a fractal regime.

After this step we went to calculate the largest Lyapunov exponent. The following results were obtained: $\lambda_E = 0.055 \pm 0.032$ in the case of Cal1-309, $\lambda_E = 0.035 \pm 0.031$ for Cal1-400, $\lambda_E = 0.052 \pm 0.030$ in the case of Cuc1-309, $\lambda_E = 0.050 \pm 0.030$ in the case of Cuc1-400, $\lambda_E = 0.063 \pm 0.033$ in the case of Raf1-309 and $\lambda_E = 0.066 \pm 0.031$ in the case of Raf1-400.

All the estimated values of largest Lyapunov exponents resulted to be positive. The conclusion is that Glow Curve might exhibit a chaotic regime. Analysis of Lyapunov spectrum confirmed that we are in presence of a dissipative system.

Analysis of surrogate data was performed using shuffled data. For Cal1-309 the autocorrelation function gave value $\tau = 0.62$ and $\lambda_E = 0.105 \pm 0.052$ for the largest Lyapunov exponent. For Cal1-400 we had $\tau = 0.73$ and $\lambda_E = 0.101 \pm 0.043$. For Cuc1-309 we had $\tau = 0.66$ and $\lambda_E = 0.090 \pm 0.050$. For Cuc1-400 we had $\tau = 0.63$ and $\lambda_E = 0.073 \pm 0.041$. Finally, for Raf1-309 we obtained $\tau = 0.73$ and $\lambda_E = 0.147 \pm 0.061$, and for Raf1-400 it resulted $\tau = 0.67$ and $\lambda_E = 0.140 \pm 0.044$.

Comparing normal and shuffled data, so marked differences appear that the null hypothesis was rejected and our conclusion was that Glow Curves in thermoluminescence exhibit chaotic dynamics.

The final step was to perform Recurrence Quantification Analysis (RQA) for the given samples. This is a kind of analysis that will enable us to explore for the first time the inner structure of a given Glow Curve signal.

For the analysis a Radius R=16 and a Line L=2 were employed. Euclidean distance and rescaling mean were used.

Sample Cal1-309:

%Rec=1.226 , %Det=21.429 , %Lam = 23.443 , T.T = 2.909 , Ratio = 17.485 , En =2.372 , Mline =14 , Trend =-13.705.
Sample Call-400:
%Rec=3.202 , %Det=40.428 , %Lam = 56.896 , T.T = 3.338 , Ratio = 12.625, En =3.249 , Mline =26 , Trend =- 28.407.
Sample Cuc1-309:
%Rec=4.002 , %Det=44.604 , %Lam = 64.386 , T.T = 3.739 , Ratio = 11.155, En =3.414 , Mline =31 , Trend =- 41.165.
Sample Cuc1-400:
%Rec=5.224 , %Det=63.891, %Lam = 78.451 , T.T = 5.779 , Ratio = 12.229, En =4.024, Mline =113 , Trend =- 36.583.
Sample Raf1-309:
%Rec=5.818 , %Det=68.100, %Lam = 79.558 , T.T = 5.264 , Ratio = 11.704, En =3.863, Mline =84, Trend =- 57.149.
Sample Raf1-400:
%Rec=7.165 , %Det=77.563, %Lam = 86.979 , T.T = 7.171 , Ratio = 10.325, En =4.000, Mline =177, Trend =- 43.165.

As example, Figure 5 gives the recurrence plot of a sample.

We conclude that we have recurrent signals with recurrences enclosed in the range of (1.2-7.1)%.

The Determinism of Glow Curves is very moderate , resulting in the range (21.4-77.5)%.

Regimes of laminarity are enclosed in the range (23.4-86.9)% with times for states of laminarity ranging from 2.9 to 7.1 .Glow Curves are dynamic processes supported from intermittency. This is to say that we have rather frequent regimes of transition chaos-chaos , and chaos-order. Entropy is very high , ranging from 2.3 to 4.00 bits/°K . The ratio remains rather constant , about 10-12 with the only exception of Call-£09 where raises to 17.4 , and , finally, Mline results variable from 14 to 177. The inverse of MaxLine values gives a measure of the divergence of the dynamics of the samples.

The conclusion seems to be that we are in presence of a signal showing an high level of complexity marked from profound variability and variedness.

Finally, it was conducted analysis of fractal dimension by calculation of Hurst exponent.

It resulted that

Call-309 : H=1.00510 ± 0.00899 , $r^2$ = 0.984
Call-400 : H=1.05389 ± 0.00867 , $r^2$ = 0.994
Cuc1-309 : H=1.03956 ± 0.00686 , $r^2$ = 0.995
Cuc1-400 : H=1.08117 ± 0.00702 , $r^2$ = 0.997
Raf1-309 : H=1.05047 ± 0.00644 , $r^2$ = 0.997
Raf1-400 : H=1.06759 ± 0.00511 , $r^2$ = 0.998

In conclusion, in Glow Curves of thermoluminescence we have also fractal regimes (fractional Brownian motion) signed from persistence , that is to say long range correlations during electrons release with variation of the temperature.

*Conclusions.* Thermoluminescence is a process that follows quantum mechanics. In fact, it may be explained by using the band theory [13,14]:

The ionization of atoms, due to radiations, releases electrons from the valence band. Consequently , some holes are formed, and the electrons are captured by the traps formed by the imperfections or impurity atoms, or atoms of doping, in the forbidden band of the crystal. They are then in a metastable state. According to the type of crystalline material they may remain from a very short time to thousands of years. Heat energy applied to the crystal allows the electrons to leave the traps. They, then, come back in the valence band while emitting some light and producing the thermoluminescence phenomenon. Such quantum processes may be characterized from the presence of quantum entanglement as previously studied and analysed in detail , under the theoretical and experimental profiles in various papers given in [14,15] and references therein. As previously outlined, in thermoluminescence (TL) dating , a TL curve is obtained when a solid sample , taken as example from a pottery, is heated after that it has remained exposed to natural irradiations ($\alpha, \beta, \gamma$) for a certain number of years. There are electrons which have been ionized by nuclear radiation and which have diffused into the vicinity of a defect in the lattice that is attractive to electrons, such as a negative-ion vacancy, and have become trapped there. The nuclear radiation is from radioelements in the sample and in its surroundings; but there is also a small contribution from cosmic rays. Still , it is known that low energy photon pairs from atomic radioactive cascades are entangled [14,15]. Generally speaking, also in electron-positron annihilation the two emitted gamma rays are entangled. Still, in bremsstrahlung one electron produces several photons instantaneously and such photons are entangled according to Quantum Mechanics ,and discussed in detail still in [14] and [15] , in various papers, from authors R. Desbrandes, and D. L. Van Gent. In addition , in [14,15] , it is still evidenced that in thermoluminescence the quantum phenomenon of entanglement swapping is possible. These authors correctly outline that entanglement can be swapped between two particles and two other particles : entangled particles, such as electrons, can be stored in ion traps or impurities within thermoluminescent crystal lattices and remain isolated from environmental decoherence effects in the traps for long time. Following thermal heating , the electrons are then forced to leave these traps and then drop down to their ground state energies in the crystal lattice . An entangled electron dropping out of its ion trap will go through spin transitions which affect its corresponding entangled electron by reason of spin conservation laws. In this manner it becomes favourable for the other entangled electron to exit its trap as a result, emitting some light while dropping to ground state. For detailed explanations , theoretical treatments , experimental and developments at the level of application, see still ref .[14,15] . Thus, in conclusion, by thermoluminescence we are in presence of a well definite quantum mechanical regime. On the other hand our results evidence that, when the sample is subjected to thermal heating, a chaotic regime arises in the dynamics of emission of light intensity. Chaos arises when we are in presence of decoherence effects and it is just such combination of quantum mechanical framework from one hand with chaos counterpart on the other hand that renders the interesting features of the results that we have obtained in the present paper. In fact, rather recently , the possible links between chaos and entanglement have started to be discussed [16].

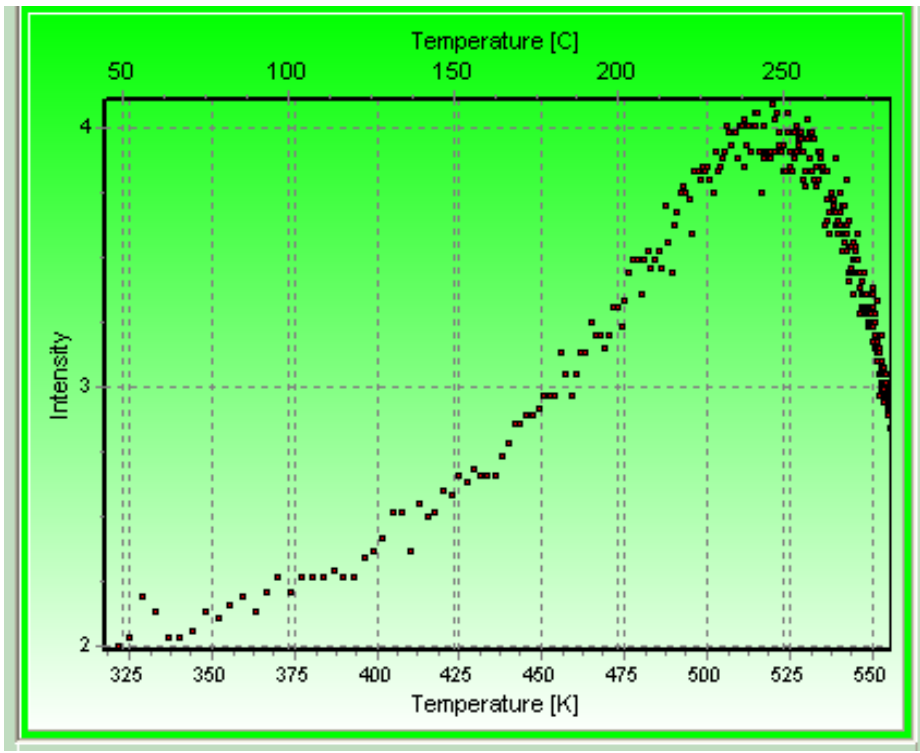

Figure 1: Cuc1-309 -Glow Curve

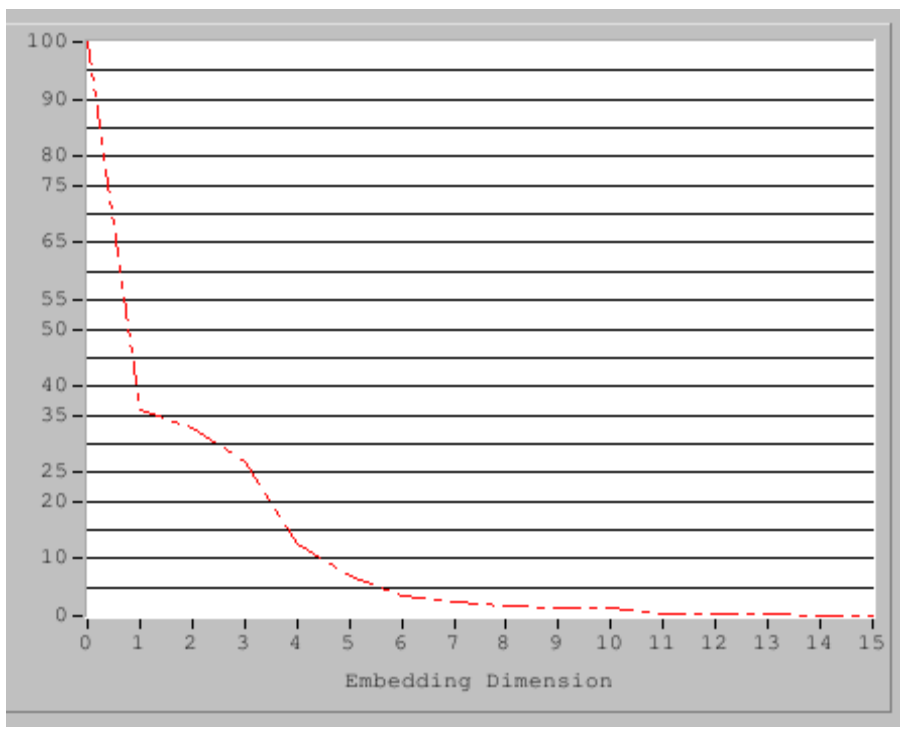

Figure 2: Analysis of False Nearest Neighbors (Cuc1-309)

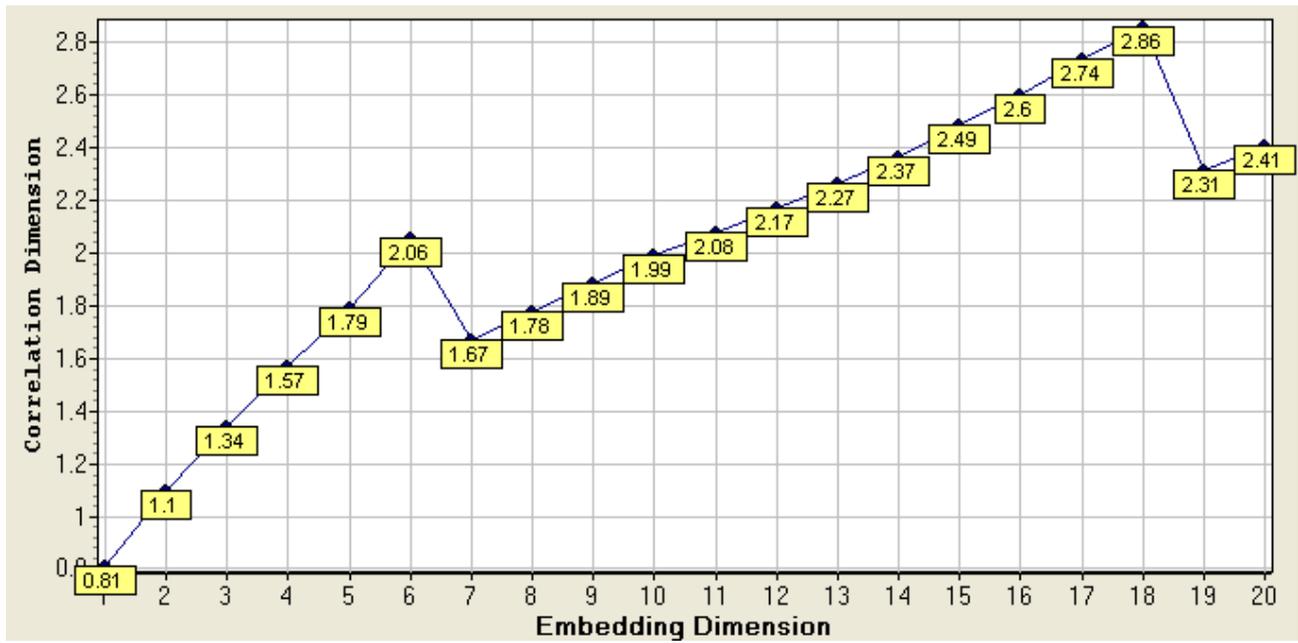

Figure 3: Estimation of attractor dimension (Cuc1-309)

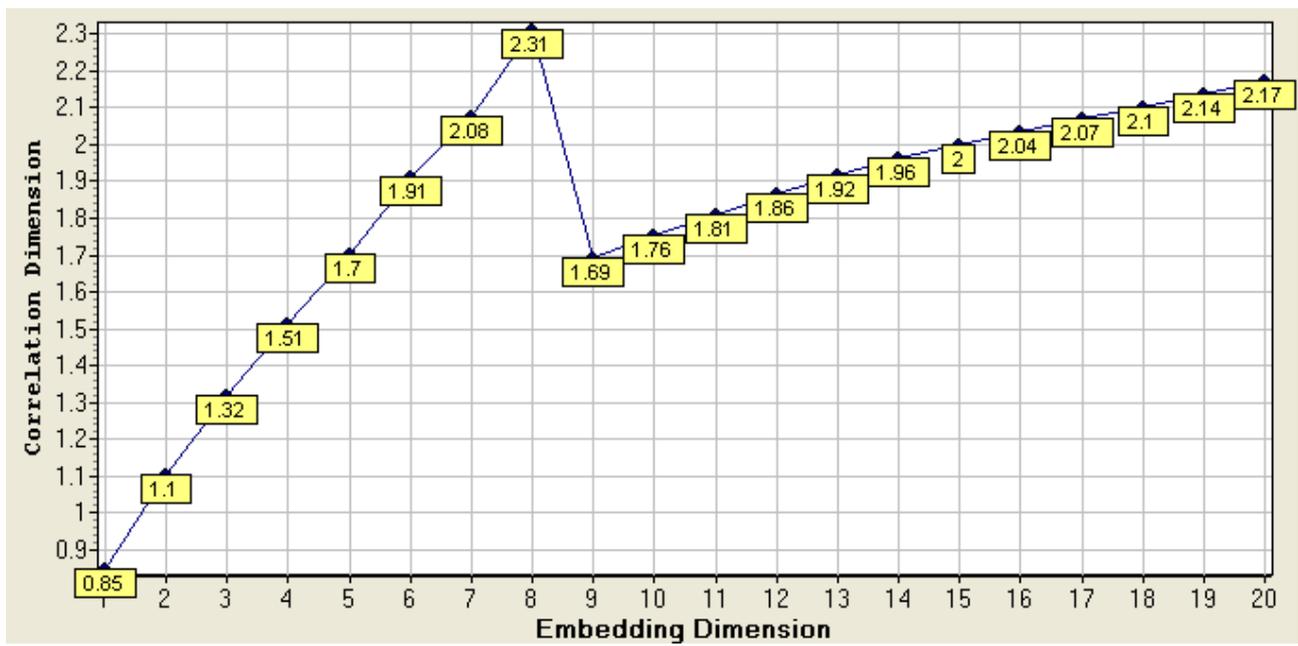

Figure 4: Estimation of attractor dimension (Cuc1-400)

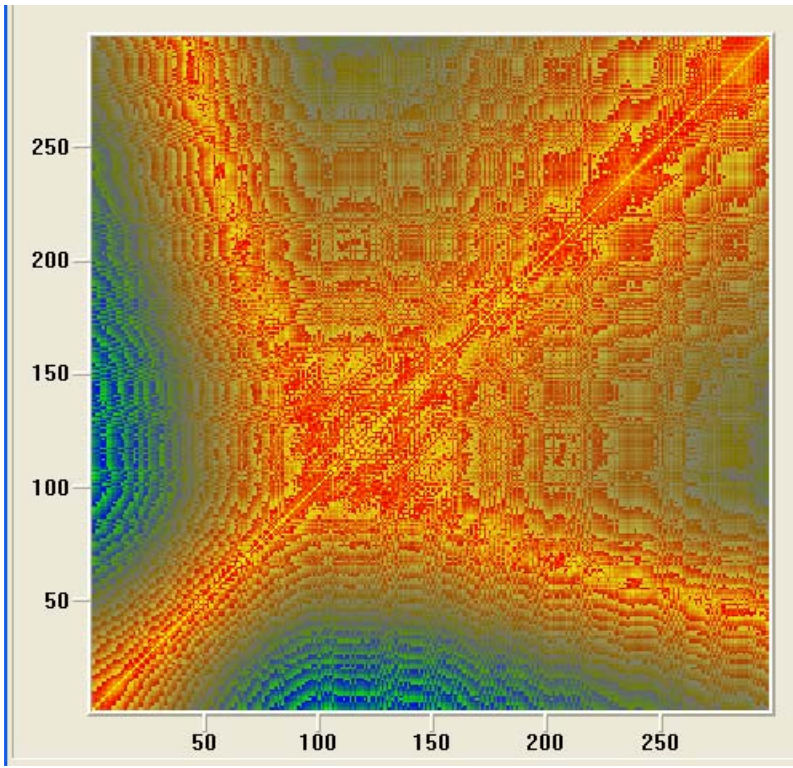
Figure 5:Recurrences plot of Cuc1-309